\documentstyle[psfig,12pt]{article}
\def\TL{\hfil$\displaystyle{##}$}
\def\TR{$\displaystyle{{}##}$\hfil}
\def\TC{\hfil$\displaystyle{##}$\hfil}
\def\TT{\hbox{##}}
\def\seqalign#1#2{\vcenter{\openup1\jot
  \halign{\strut #1\cr #2 \cr}}}

\def\comment#1{}
\def\fixit#1{}

\def\tf#1#2{{\textstyle{#1 \over #2}}}

\def\mop#1{\mathop{\rm #1}\nolimits}

\def\Vol{\mop{Vol}}

\def\diag{\mop{diag}}
\def\tr{\mop{tr}}
\def\Disc{\mop{Disc}}

\def\overleftrightarrow#1{\vbox{\ialign{##\crcr
     $\leftrightarrow$\crcr\noalign{\kern-0pt\nointerlineskip}
     $\hfil\displaystyle{#1}\hfil$\crcr}}}

\def\lsim{\mathrel{\mathstrut\smash{\ooalign{\raise2.5pt\hbox{$<$}\cr\lower2.5pt\hbox{$\sim$}}}}}
\def\gsim{\mathrel{\mathstrut\smash{\ooalign{\raise2.5pt\hbox{$>$}\cr\lower2.5pt\hbox{$\sim$}}}}}

\def\sqr#1#2{{\vcenter{\vbox{\hrule height.#2pt
         \hbox{\vrule width.#2pt height#1pt \kern#1pt
            \vrule width.#2pt}
         \hrule height.#2pt}}}}
\def\square{\mathop{\mathchoice\sqr56\sqr56\sqr{3.75}4\sqr34\,}\nolimits}

\def\href#1#2{#2}  

\def\lbldef#1#2{\expandafter\gdef\csname #1\endcsname {#2}}
\def\eqn#1#2{\lbldef{#1}{(\ref{#1})}%
\begin{equation} #2 \label{#1} \end{equation}}
\def\eqalign#1{\vcenter{\openup1\jot
    \halign{\strut\span\TL & \span\TR\cr #1 \cr
   }}}
\def\eno#1{(\ref{#1})}

\def\comment#1{  \begin{raggedright}{\tt [#1]}\end{raggedright}}
\begin{document}
\baselineskip=15.5pt
\pagestyle{plain}
\setcounter{page}{1}
\renewcommand{\thefootnote}{\fnsymbol{footnote}}

\begin{titlepage}

\begin{flushright}
CERN-TH/99-189 \\
HUTP-99/A029 \\
MIT-CTP-2877 \\
USC-99/03 \\
hep-th/9906194
\end{flushright}
\vfil

\begin{center}
{\huge Continuous distributions of D3-branes} \\[8pt]
{\huge and gauged supergravity}
\end{center}

\vfil
\begin{center}
{\large D.Z. Freedman$^{1}$, S.S. Gubser$^{2}$, K. Pilch$^{3}$, and 
N.P. Warner$^{3,4}$}
\end{center}

$$\seqalign{\span\TL & \span\TT}{
^1 & Department of Mathematics and Center for Theoretical Physics,
  \cr\noalign{\vskip-1.5\jot}
   & Massachusetts Institute of Technology, Cambridge, MA  02139-4307, USA 
  \cr\noalign{\vskip1\jot}
^2 & Lyman Laboratory of Physics, Harvard University, Cambridge, MA
02138, USA 
  \cr\noalign{\vskip1\jot}
^3 & Department of Physics and Astronomy, University of Southern California,
  \cr\noalign{\vskip-1.5\jot}
   & Los Angeles, CA 90089-0484, USA 
  \cr\noalign{\vskip1\jot}
^4 & Theory Division, CERN, CH-1211 Geneva 23, Switzerland
}$$
\vfil

\begin{center}
{\large Abstract}
\end{center}

\noindent
States on the Coulomb branch of ${\cal N}=4$ super-Yang-Mills theory are
studied from the point of view of gauged supergravity in five dimensions.
These supersymmetric solutions provide examples of consistent truncation
from type IIB supergravity in ten dimensions. A mass
gap for states created by local operators and perfect screening for
external quarks arise in the supergravity approximation.  We offer an
interpretation of these surprising features in terms of ensembles of brane
distributions.

\vfil
\begin{flushleft}
June 1999
\end{flushleft}
\end{titlepage}
\newpage

\def\IR{\relax{\rm I\kern-.18em R}}
\def\cN{{\cal N}}
\def\del{\partial}
\def\IR{\relax{\rm I\kern-.18em R}}
\def\cN{{\cal N}}
\def\ie{{\it i.e.} }
\def\bfone{\relax{\rm 1\kern-.35em 1}}
\def\inbar{\vrule height1.5ex width.4pt depth0pt}
\def\LG{Lan\-dau-Ginz\-burg\ }
\def\IC{\relax\,\hbox{$\inbar\kern-.3em{\rm C}$}}
\def\ID{\relax{\rm I\kern-.18em D}}
\def\IF{\relax{\rm I\kern-.18em F}}
\def\IH{\relax{\rm I\kern-.18em H}}
\def\II{\relax{\rm I\kern-.17em I}}
\def\IN{\relax{\rm I\kern-.18em N}}
\def\IP{\relax{\rm I\kern-.18em P}}
\def\IQ{\relax\,\hbox{$\inbar\kern-.3em{\rm Q}$}}
\def\us#1{\underline{#1}}
\def\IR{\relax{\rm I\kern-.18em R}}
\font\cmss=cmss10 \font\cmsss=cmss10 at 7pt
\def\ZZ{\relax\ifmmode\mathchoice
{\hbox{\cmss Z\kern-.4em Z}}{\hbox{\cmss Z\kern-.4em Z}}
{\lower.9pt\hbox{\cmsss Z\kern-.4em Z}}
{\lower1.2pt\hbox{\cmsss Z\kern-.4em Z}}\else{\cmss Z\kern-.4em
Z}\fi}
\def\a{\alpha} \def\b{\beta} \def\d{\delta}
\def\e{\epsilon} \def\c{\gamma}
\def\G{\Gamma} \def\l{\lambda}
\def\L{\Lambda} \def\s{\sigma}
\def\cA{{\cal A}} \def\cB{{\cal B}}
\def\cC{{\cal C}} \def\cD{{\cal D}}
\def\cF{{\cal F}} \def\cG{{\cal G}}
\def\cH{{\cal H}} \def\cI{{\cal I}}
\def\cJ{{\cal J}} \def\cK{{\cal K}}
\def\cL{{\cal L}} \def\cM{{\cal M}}
\def\cN{{\cal N}} \def\cO{{\cal O}}
\def\cP{{\cal P}} \def\cQ{{\cal Q}}
\def\cR{{\cal R}} \def\cS{{\cal S}}
\def\cU{{\cal U}} \def\cV{{\cal V}}
\def\om{\omega}
\def\vt{\vartheta}
\def\nup#1({Nucl.\ Phys.\ $\us {B#1}$\ (}
\def\plt#1({Phys.\ Lett.\ $\us  {B#1}$\ (}
\def\cmp#1({Comm.\ Math.\ Phys.\ $\us  {#1}$\ (}
\def\prp#1({Phys.\ Rep.\ $\us  {#1}$\ (}
\def\prl#1({Phys.\ Rev.\ Lett.\ $\us  {#1}$\ (}
\def\prv#1({Phys.\ Rev.\ $\us  {#1}$\ (}
\def\mpl#1({Mod.\ Phys.\ Let.\ $\us  {A#1}$\ (}
\def\ijmp#1({Int.\ J.\ Mod.\ Phys.\ $\us{A#1}$\ (}
\def\jag#1({Jour.\ Alg.\ Geom.\ $\us {#1}$\ (}
\def\tit#1|{{\it #1},\ }
\def\Coeff#1#2{{#1\over #2}}
\def\Coe#1.#2.{{#1\over #2}}
\def\coeff#1#2{\relax{\textstyle {#1 \over #2}}\displaystyle}
\def\coe#1.#2.{\relax{\textstyle {#1 \over #2}}\displaystyle}
\def\half{{1 \over 2}}
\def\shalf{\relax{\textstyle {1 \over 2}}\displaystyle}
\def\dag#1{#1\!\!\!/\,\,\,}
\def\del{\partial}
\def\nex#1{$N\!=\!#1$}
\def\dex#1{$d\!=\!#1$}
\def\cex#1{$c\!=\!#1$}
\def\doubref#1#2{\refs{{#1},{#2}}}
\def\multref#1#2#3{\nrf{#1#2#3}\refs{#1{--}#3}}
\def\br{\hfill\break}
\def\wT{\widetilde T}
\def\lsw{\lambda_{\rm SW}}
\def\tE{\widetilde E}
\def\barE{\overline{E}}
\def\hE{\widehat E}
\def\ttau{\widetilde \tau}
\def\tq{\widetilde q}
\def\Imt{{\rm Im}\tau}

\section{Introduction}

The AdS/CFT correspondence \cite{juanAdS,gkPol,witHolOne} has been
primarily studied in the conformal vacuum of ${\cal N}=4$ super-Yang-Mills
theory.  However, it also includes other states in the Hilbert space of the
theory; these correspond to certain solutions of the supergravity field
equations in which the bulk space-time geometry approaches $AdS_5$ near the
boundary, but differs from $AdS_5$ in the interior.  Sometimes a simpler
picture of states in the gauge theory emerges from the ten-dimensional
geometry.  For instance, two equal clusters of coincident D3-branes
separated by a distance $\ell$ correspond to the vacuum state of the gauge
theory where the $SU(N)$ gauge symmetry has been broken to $SU(N/2) \times
SU(N/2)$ by scalar vacuum expectation values (VEV's).  This configuration
has been studied in \cite{JMNW,9905104}.  More generally, one could
consider any distribution of the $N$ D3-branes in the six transverse
dimensions.  These configurations preserve sixteen supersymmetries, as
appropriate since the Poincar\'e supersymmetries of the gauge theory are
maintained but superconformal invariance is broken by the Higgsing.  The
space of possible distributions is precisely the moduli space ${\rm Sym}^N
\, {\bf R}^6$ of the gauge theory.  It is known as the Coulomb branch
because the gauge bosons which remain massless mediate long-range Coulomb
interactions.

At the origin of moduli space, where all the branes are coincident, the
near-horizon geometry is $AdS_5 \times S^5$.  Each factor has a radius of
curvature $L$ given by
  \eqn{LDef}{
   L^4 = {\kappa_{10} N \over 2 \pi^{5/2}} \ ,
  }
 where $\kappa_{10}$ is the ten-dimensional gravitational constant.  If the
branes are not coincident, but the average distance $\ell$ between them is
much less than $L$, then the geometry will still have a near-horizon region
which is asymptotically $AdS_5 \times S^5$.  From the five-dimensional
perspective, the deviations from this limiting geometry arise through
non-zero background values for scalars in the supergravity theory.  At
linear order these background values are solutions of the free
wave-equations for the scalars with regular behavior near the boundary of
$AdS_5$, and so we recover the usual picture of states in the gauge theory
in AdS/CFT.  Given a particular vacuum state, specified by a distribution
of branes in ten dimensions or an asymptotically $AdS_5$ geometry with
scalar profiles in five dimensions, it is natural to ask what predictions
the correspondence makes regarding Green's functions and Wilson loops.

From the point of view of supergravity, the two-center solution is
complicated because infinitely many scalar fields are involved in the five
dimensional description.  The present paper is therefore concerned with
states on the Coulomb branch that are simple from the point of view of
supergravity: they will involve only the scalars in the massless ${\cal
N}=8$, five-dimensional supergraviton multiplet.  More specifically, we 
will investigate geometries involving profiles for the supergravity
modes dual to the operators
  \eqn{XXModes}{
   \tr X_{(i} X_{j)} = \left( \delta_i^k \delta_j^l - 
    \tf{1}{6} \delta_{ij} \delta^{kl} \right) \tr X_k X_l \ .
  }
 These operators and their dual fields in supergravity transform in the
$20'$ of $SO(6)$.

All the geometries we consider preserve sixteen supercharges, and this
allows us to reduce the field equations to a first-order system.  The
geometries naturally separate into five universality classes, identified
according to the asymptotic behavior far from the boundary of $AdS_5$.
There is a privileged member in each class which preserves $SO(n) \times
SO(6-n)$ of the $SO(6)$ local gauge symmetry for $n=1,2,3,4,5$ (as usual
$SO(1)$ is the trivial group).  We identify the distribution of D3-branes
in ten dimensions which leads to each of the privileged geometries: in each
case the distribution is a $n$-dimensional ball.  Next, we investigate the
behavior of two-point correlators and Wilson loops.  Surprisingly, we find
a mass gap in the two-point correlator for $n=2$ and a completely discrete
spectrum for $n>2$.
Also, Wilson loops exhibit perfect screening for $n \geq 2$ for
quark-anti-quark separations larger than the inverse mass gap.  We suggest
a tentative interpretation of these results in terms of an average over
positions of branes within the distribution.

This study is
an outgrowth of the work of \cite{fgpw}, in which it was found that there
are soliton solutions of the supergravity theory which preserve ${\cal
N}=1$ Poincar\'e supersymmetry.  The scalar fields in these ${\cal N}=1$
flows lie in two-dimensional submanifolds of the $42$-dimensional scalar
coset $E_{6(6)}/USp(8)$, of which one field is a component of the $20'$ and
the other of the $10 \oplus \overline{10}$ representation.  The $n=2$ and
$n=4$ geometries considered below correspond to special solutions of the
flow equations of \cite{fgpw} in which the $10 \oplus \overline{10}$
component vanishes and supersymmetry is enhanced to ${\cal N}=4$.

\section{Supergravity solutions}
\label{SUGRASolns}

Maximal ${\cal N}=8$ gauged supergravity in four dimensions is a consistent
truncation of eleven-dimensional supergravity compactified on $S^7$ (see
\cite{BdWHN} and references therein), and the same has recently been
demonstrated for the maximal gauged supergravity theory in seven
dimensions \cite{stny}. 
There is little doubt that maximal
${\cal N}=8$ gauged supergravity in five dimensions is likewise a
consistent truncation of ten-dimensional type~IIB supergravity on $S^5$,
although a formal proof has not been given.  Consistent truncation means
that fields of the parent theory and its truncation are related by an
Ansatz such that any solution of the equations of motion of the truncated
theory lifts unambiguously to a solution of the parent theory.

Gauged ${\cal N}=8$ supergravity in five dimensions \cite{GRW,PPvN,GRWb}
involves $42$ scalars parametrizing the coset $E_{6(6)}/USp(8)$.  An
important ingredient in consistent truncation arguments is a map that takes
any element of the coset to a particular deformed metric on $S^5$.  The
identity element is associated to the round $S^5$.  The general
form of the map was essentially given in \cite{kpw}, in terms of the
scalar $27$-bein, ${\cal V}$, of $E_{6(6)}/USp(8)$.  The
resulting ten-dimensional metric $d\hat{s}^2$ has the form of a warped
product:
  \eqn{WarpedProduct}{ 
   d\hat{s}^2 = \Delta^{-2/3} ds_M^2 + ds_K^2 \ , 
  } 
 where $ds_M^2$ is the metric on the five noncompact coordinates and
$ds_K^2$ is the metric on the deformed $S^5$.  The warp factor $\Delta$
depends on the $S^5$ coordinates, and it is roughly the local dilation of
the volume element of $S^5$.  There can be $M$-dependence in $ds_K^2$ but
not vice versa.

The group $E_{6(6)}$ contains $SL(6,{\bf R}) \times SL(2,{\bf R})$ as a
maximal subgroup.  The $20'$ of scalars which we want to consider
parametrizes the coset $SL(6,{\bf R})/SO(6)$.  Choosing a representative $S
\in SL(6,{\bf R})$ for a specified element of the coset, we can form the
symmetric matrix $M = S S^T$.  The theory depends only on this combination,
and the $SO(6)$ symmetry acts on it by conjugation.  So we may take $M$ to
be diagonal:
  \eqn{MDef}{
   M = \diag\{ e^{2\beta_1}, e^{2\beta_2}, e^{2\beta_3}, e^{2\beta_4}, 
    e^{2\beta_5}, e^{2\beta_6} \} \ .
  }
The $\beta_i$ sum to zero, and we take the following
convenient orthonormal parametrization:
  \eqn{betaDef}{
   \pmatrix{ \beta_1 \cr \beta_2 \cr \beta_3 \cr \beta_4 \cr \beta_5 \cr 
     \beta_6 } = 
    \pmatrix{ 1/\sqrt{2} & 1/\sqrt{2} & 1/\sqrt{2} & 0 & 1/\sqrt{6} \cr
              1/\sqrt{2} & -1/\sqrt{2} & -1/\sqrt{2} & 0 & 1/\sqrt{6} \cr
              -1/\sqrt{2} & -1/\sqrt{2} & 1/\sqrt{2} & 0 & 1/\sqrt{6} \cr
              -1/\sqrt{2} & 1/\sqrt{2} & -1/\sqrt{2} & 0 & 1/\sqrt{6} \cr
              0 & 0 & 0 & 1 & -\sqrt{2/3} \cr
              0 & 0 & 0 & -1 & -\sqrt{2/3} }
    \pmatrix{ \alpha_1 \cr \alpha_2 \cr \alpha_3 \cr \alpha_4 \cr 
      \alpha_5 }
  }
 The relevant part of the ${\cal N}=8$ lagrangian \cite{GRWb}, in
$+$$-$$-$$-$ signature, is
  \eqn{NEightL}{
   {\cal L} = -{1 \over 4} R + \sum_{i=1}^5 {1 \over 2} 
    (\partial \alpha_i)^2 - P
  }
 where
  \eqn{VDef}{
   P = -{g^2 \over 32} \left[ (\tr M)^2 - 2 \tr M^2 \right] \ .
  }
 In analogy with the results of \cite{fgpw} it is possible to show that
  \eqn{WRel}{
   P = {g^2 \over 8} \sum_{i=1}^5 
    \left( {\partial W \over \partial\alpha_i} \right)^2 - 
    {g^2 \over 3} W^2 \qquad \hbox{where} \quad W = -{1 \over 4} \tr M \ .
  }
 It is also straightforward to show that the tensor $W_{ab}$ which enters
the gravitino transformations as
  \eqn{PsiTransform}{
   \delta \psi_{\mu a} = {\cal D}_\mu \epsilon_a - {g \over 6} 
    W_{ab} \gamma_\mu \epsilon^b
  }
 has the form $W_{ab} = W \delta_{ab}$.  The Killing spinor conditions
 are $\delta \psi_{\mu a} =0$ and $\delta \chi_{abc}=0$, where
$\chi_{abc}=0$ are the $48$ spin $1/2$ fields of the theory. It can be shown
that sixteen supercharges are preserved if and only if
  \eqn{SUSY}{
   {d \alpha_i \over d\rho} =
    {g \over 2} {\partial W \over \partial\alpha_i}
   \qquad \hbox{and} \qquad
   {d A \over d\rho} = -{g \over 3} W \ ,
  }
where $\rho$ is the radial coordinate of a metric of the form
\eqn{MetricAnsatz}{
   ds^2 = e^{2 A(\rho)} dx_\mu^2 - d\rho^2 \ .
  }

There are no extrema of $W$ except for the global maximum when all the
$\beta_a$ are $0$.\footnote{There is however an extremum of $V$ at 
$\beta_a = -{\log 3 \over 12} (-5,1,1,1,1,1)$.  
This is the known unstable $SO(5)$
invariant critical point of the theory \cite{GRW,dz}. }
All flows have some
$\beta_a \to \pm \infty$ in finite or infinite $\rho$.  Asymptotically they
must approach a fixed direction: $\beta_a \to \gamma_a \mu$ where
$\gamma_a$ is a fixed vector, with $\gamma^2 = 2$ so that $\mu$ is
canonically normalized.  The sign of $\gamma_a$ matters because we take
$\mu \to +\infty$ far from the boundary of $AdS_5$.  It is straightforward
to verify that the possible $\gamma_a$ are those listed in
Table~\ref{tableA}.

For each $\gamma_a$, there is a privileged flow determined by the condition
that $\beta_a = \gamma_a \mu$ exactly all along the flow, rather than just
asymptotically.  This condition leaves just one parameter of freedom to
determine the flow: a quantity $\ell^2$ which controls the size of $\mu$ near
the boundary of $AdS_5$ and is proportional to $\langle {\cal O}_{20'}
\rangle$, where ${\cal O}_{20'}$ is the operator dual to $\mu$.  The
symmetry groups preserved by the privileged flows are also listed in
Table~\ref{tableA}.  Each of them can be lifted unambiguously to a
ten-dimensional geometry, which in each case can be written in the form
  \eqn{TenForm}{\eqalign{
   ds^2 &= {1 \over \sqrt{H}} \left( dt^2 - dx_1^2 - dx_2^2 - dx_3^2 \right) -
    \sqrt{H} \sum_{i=1}^6 dy_i^2  \cr
   H &= \int_{|\vec{w}| \leq \ell} 
    d^n w \, \sigma(\vec{w}) {L^4 \over |\vec{y}-\vec{w}|^4} \ .
  }}
 The $n$-dimensional integral is over a ball of radius $\ell$ in $n$ of the
six dimensions transverse to the D3-branes.  The distribution of branes,
$\sigma(\vec{w})$, depends only on $|\vec{w}|$ and is normalized
to $1$.  The various $\sigma_n$ as functions of $w = |\vec{w}|$ are listed
in Table~\ref{tableA}.  It is amusing to note that if one starts with the
uniform disk of branes specified by $\sigma_2$ and compresses it to a line
segment by projecting the position of each brane perpendicularly onto one
axis, the result is the distribution $\sigma_1$.  The analogous projection
relations obtain between $\sigma_n$ and $\sigma_{n-1}$ for $n=3,4,5$.  The
distribution $\sigma_5$ has the form
  \eqn{nFiveSigma}{
   \sigma_5(w) = {1 \over \pi^3 \ell^2} \left(
    -{1 \over 2} {1 \over (\ell^2-w^2)^{3/2}} \theta(\ell^2-w^2) + 
     {1 \over \sqrt{\ell^2-w^2}} \delta(\ell^2-w^2) \right) \ .
  }
 Unlike all the other $\sigma_n$, $\sigma_5$ is not uniformly positive.
This is a deep pathology which leads us to conclude that this geometry is
unphysical.  It cannot even be interpreted in terms of anti-D3-branes:
negative ``charge'' in $\sigma_5$ indicates an object of negative tension
as well as opposite Ramond-Ramond charge to the D3-brane.  The only such
objects in string theory are orientifold planes, but to make up the
$\sigma_5$ in \nFiveSigma\ one would require infinitely many O3-planes,
which again seems senseless.  It is curious that the $n=5$ case has such
pathological ten-dimensional origins: in five dimensions its naked
singularity is of much the same type as for $n<5$.  
  \begin{table}
  \begin{center}
  \begin{tabular}{llllc}
   $ n $&$ \hfil \gamma_a \hfil $& symmetry &$ \hfil \sigma \hfil $& 2pt fnct  
    \\[3pt] \hline \\[-7pt] 
   $ 1 $&$ \displaystyle{{1 \over \sqrt{15}} (1,1,1,1,1,-5)} $&$ SO(5) $&$
    \displaystyle{{2 \over \pi \ell^2} \sqrt{\ell^2-w^2}} $& continuum  \\[9pt]
   $ 2 $&$ \displaystyle{{1 \over \sqrt{6}} (1,1,1,1,-2,-2)} $&$ 
    SO(4) \times SO(2) $&$
    \displaystyle{{1 \over \pi\ell^2} \theta(\ell^2-w^2)} $& gapped  \\[9pt]
   $ 3 $&$ \displaystyle{{1 \over \sqrt{3}} (1,1,1,-1,-1,-1)} $&$
    SO(3) \times SO(3) $&$
    \displaystyle{{1 \over \pi^2 \ell^2} {1 \over \sqrt{\ell^2-w^2}}}
     $& discrete  \\[9pt]
   $ 4 $&$ \displaystyle{{1 \over \sqrt{6}} (2,2,-1,-1,-1,-1)} $&$
    SO(4) \times SO(2) $&$
     \displaystyle{{1 \over \pi^2 \ell^2} \delta(\ell^2-w^2)} $& discrete  \\[9pt]
   $ 5 $&$ \displaystyle{{1 \over \sqrt{15}} (5,-1,-1,-1,-1,-1)} $&$ SO(5) $&
    {\rm Eqn.} \eno{nFiveSigma} & discrete
  \end{tabular}
  \end{center}
 \caption{A summary of the privileged ${\cal N}=4$ geometries and their
properties.  It is helpful to note that the distributions $\sigma$ vanish
by definition for $w>\ell$.}\label{tableA}
  \end{table}

The distribution $\sigma_2$ was considered previously in \cite{klt} in
connection with a zero temperature, zero angular momentum limit of a
spinning D3-brane metric with angular momentum in a single plane
perpendicular to the branes.  As shown in \cite{Myers,CvGu1,PopeEtAl}, the
Kaluza-Klein reduction of the spinning brane geometry to five dimensions
involves only the fields of the gauged supergravity multiplet, and in fact
it is a non-extremal R-charged black hole of the type discussed in
\cite{bcs}.  Indeed the five-dimensional geometry corresponding to $n=2$
can be shown to be precisely the extremal limit of this black hole geometry
where the mass approaches the charge from above: $M \to Q^+$ in appropriate
five-dimensional units.  Amusingly, the $n=4$ geometry is precisely the $M
\to Q^-$ limit of R-charged black holes whose mass is less than their
charge.  These black holes have naked timelike singularities like the
negative mass Schwarzschild solution, and they are usually deemed
unphysical.  The naked singularity remains in the $M \to Q^-$ limit, but it
is seen as a benign effect of the Kaluza-Klein reduction: the
ten-dimensional geometry has only a null singular horizon.
Geometries with the same sort of naked singularity in five dimensions have
been studied in \cite{ksDil} and also in \cite{minDil,gDil,GPPZDil} in
connection with confinement.  The well-defined ten-dimensional geometry
provides the first clear-cut evidence that such singular five-dimensional
geometries must have a role in the correspondence.  It should be noted that
the $n=4$ geometry can also be obtained as a $M \to (Q/\sqrt{2})^+$ limit
of a doubly-R-charged black hole corresponding to D3-branes with two equal
angular momenta in orthogonal planes, and that the distribution $\sigma_4$
arose in \cite{klt} in this context.

It is in principle straightforward to derive all the information in
Table~\ref{tableA} by the following strategy.  First integrate the
supersymmetry conditions \SUSY, which for our five special flows become
  \eqn{SUSYCasesstart}{\eqalign{
   &{d\mu \over d\rho} = {g \over 2} {\partial W \over \partial \mu}\,, \qquad
    {dA \over d\rho} = -{g \over 3} W\,,} }
to obtain
  \eqn{SUSYCases}{\seqalign{\span\TR \quad & \span\TR \quad & \span\TT}{
W(\mu) = -{5 \over 4} e^{{2 \over \sqrt{15}} \mu} -
         {1 \over 4} e^{-{10 \over \sqrt{15}} \mu}\,, &
	A(\mu)={1\over 2} \log\left| \,{e^{{2\over \sqrt{15}}\mu}\over
		1-e^{{12\over \sqrt{15}}\mu}} \right| 
			+\log({\ell\over L}) & for $n=1\,$,\cr
   W(\mu) = -e^{{2 \over \sqrt{6}} \mu} - 
    {1 \over 2} e^{-{4 \over \sqrt{6}} \mu}\,, &
A(\mu)={1\over 2} \log\left|\,{e^{{2\over \sqrt{6}}\mu}\over
		1-e^{{\sqrt{6}\mu}}}\right|
			+\log({\ell\over L})\,
  & for $n=2\,$,  \cr
   W(\mu) = -{3 \over 4} e^{{2 \over \sqrt{3}} \mu} - 
    {3 \over 4} e^{-{2 \over \sqrt{3}} \mu} \,, &
A(\mu)={1\over 2} \log\left|\,{e^{{2\over \sqrt{3}}\mu}\over
		1-e^{{{4\over\sqrt{6}}\mu}}}\right|
			+\log({\ell\over L})\, & for $n=3\,$,
  }}
where $\log({\ell\over L})$ is the integration constant for the first
order differential equation.
For $n=4$, $W(\mu)$ and $A(\mu)$ are  the same functions as for
$n=2$ but with $\mu \to
-\mu$; and for $n=5$,  the same as for $n=1$ but again with $\mu \to
-\mu$.  Next map the matrix $M$ to a deformed $S^5$ metric, $ds_K^2$, in
the manner described in \cite{kpw}, and use $ds_K^2$ in \WarpedProduct\ to
extract the full ten-dimensional metric.  Finally, introduce coordinates
$y_i$ transverse to the brane so that the metric assumes the form \TenForm.
The details are somewhat tedious, but it is possible to show that the
ten-dimensional metrics in their warped
product form are as follows:
  \eqn{WarpedMetrics}{\eqalign{
   n=1: & \left\{ \eqalign{
    d\hat{s}^2 &= {\zeta r^2 \over \lambda^3 L^2} 
      \left( dx_\mu^2 - {L^4 \over r^4} {dr^2 \over \lambda^6} \right) - 
     {L^2 \lambda^3 \over \zeta} 
      \left( \zeta^2 d\theta^2 + \cos^2 \theta d\Omega_4^2 \right)  \cr
    & \lambda^{12} = 1 + {\ell^2 \over r^2} \ , \quad 
      \zeta^2 = 1 + {\ell^2 \over r^2} \cos^2 \theta \ , \quad
      \Delta^{-2/3} = {\zeta \over \lambda^5}} \right.  
     \cr\noalign{\vskip2\jot}
   n=2: & \left\{ \eqalign{
    d\hat{s}^2 &= {\zeta r^2 \over L^2} 
      \left( dx_\mu^2 - {L^4 \over r^4} {dr^2 \over \lambda^6} \right) - 
     {L^2 \over \zeta} 
      \left( \zeta^2 d\theta^2 + \cos^2 \theta d\Omega_3^2 + 
       \lambda^6 \sin^2 \theta d\Omega_1^2 \right)  \cr
    & \lambda^6 = 1 + {\ell^2 \over r^2} \ , \quad 
      \zeta^2 = 1 + {\ell^2 \over r^2} \cos^2 \theta \ , \quad
      \Delta^{-2/3} = {\zeta \over \lambda^2}} \right.  
  }}
  $${\eqalign{
   n=3: & \left\{ \eqalign{
    d\hat{s}^2 &= {\zeta r^2 \lambda \over L^2} 
      \left( dx_\mu^2 - {L^4 \over r^4} {dr^2 \over \lambda^6} \right) - 
     {L^2 \over \lambda \zeta} 
      \left( \zeta^2 d\theta^2 + \cos^2 \theta d\Omega_2^2 + 
       \lambda^4 \sin^2 \theta d\widetilde{\Omega}_2^2 \right)  \cr
    & \lambda^4 = 1 + {\ell^2 \over r^2} \ , \quad 
      \zeta^2 = 1 + {\ell^2 \over r^2} \cos^2 \theta \ , \quad
      \Delta^{-2/3} = {\zeta \over \lambda}\, .} \right. 
  }}$$
 The metrics for $n=4$ and $n=5$ can be obtained from the $n=2$ and $n=1$
cases, respectively, by replacing $\ell^2 \to -\ell^2$.

\section{A two-point function}
\label{TwoPoint}

Usually the simplest two-point function to compute in supergravity is
$\langle {\cal O}_4(x) {\cal O}_4(0) \rangle$, where ${\cal O}_4 = \tr F^2
+ \ldots$ is the operator which couples to the $s$-wave dilaton.  By
$s$-wave we mean asymptotically independent of the $S^5$ coordinates.  In
ten dimensions the dilaton obeys the free wave equation, $\hat{\square}
\phi = 0$.  Solutions exist which are exactly independent of the $S^5$
coordinates (not just asymptotically), and these obey the five-dimensional
laplace equation ${\square} \phi = 0$ in the near-horizon geometry.  If
we restored the $1$ in the harmonic function $H$, then this equation would
only hold in the near-horizon region, and only in the limit where 
  \eqn{LimitL}{
   \ell \ll L \ll {1 \over \omega} \ .
  }
 Here $\omega$ is the energy of a radially infalling dilaton.  The ratio
$\omega \ell/L^2$ can be arbitrary in the limit indicated in \LimitL.  The
absorption cross-section for the dilaton is a complicated function of
$\ell$, $L$, and $\omega$, and only the leading term in small $\omega L$
and small $\ell/L$ is available via the AdS/CFT correspondence.

The properties of the five-dimensional wave equation will be most
transparent if the metric is of the form
  \eqn{zMetric}{
   ds^2 = e^{2 A(z)} \left( dt^2 - dx_1^2 - dx_2^2 - dx_3^2 - dz^2 \right) \ .
  }
 One can show that far from the boundary of $AdS_5$ one has the behavior $A
\to -\infty$ and $dA/dz \to (const.) e^{A(1-6\zeta^2)}$, where $\zeta$ is
the largest positive entry in the vector $\gamma_a$.  If $\zeta \leq
1/\sqrt{6}$, the geometry is conformal to the half of ${\bf R}^{4,1}$ where
$z>0$.  In general, curvatures are unbounded as $z \to \infty$.  If $\zeta
> 1/\sqrt{6}$, then $A \to -\infty$ at some $z = z_*$, and there is a naked
timelike singularity there.  We have $\zeta < 1/\sqrt{6}$ for $n=1$, $\zeta
= 1/\sqrt{6}$ for $n=2$, and $\zeta > 1/\sqrt{6}$ for $n>2$.

\begin{figure}
   $$\vcenter{\openup1\jot
     \halign{# & # \cr 
    $V$ & $V$  \cr\noalign{\vskip1\jot}
    \ \psfig{figure=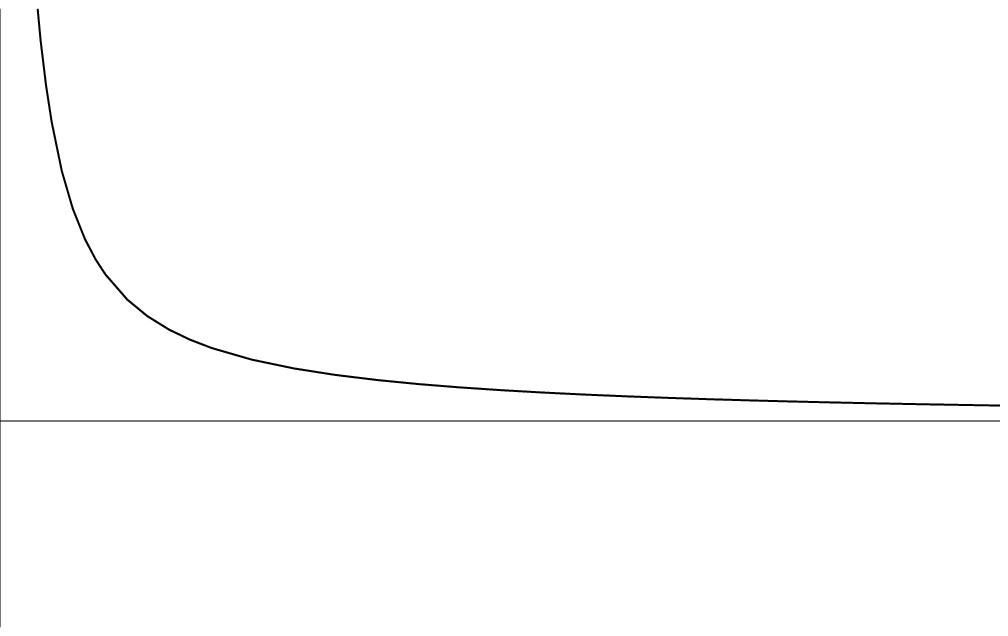,width=1.5in} \ 
     \ooalign{\raise20pt\hbox{$z$}} & 
    \ \psfig{figure=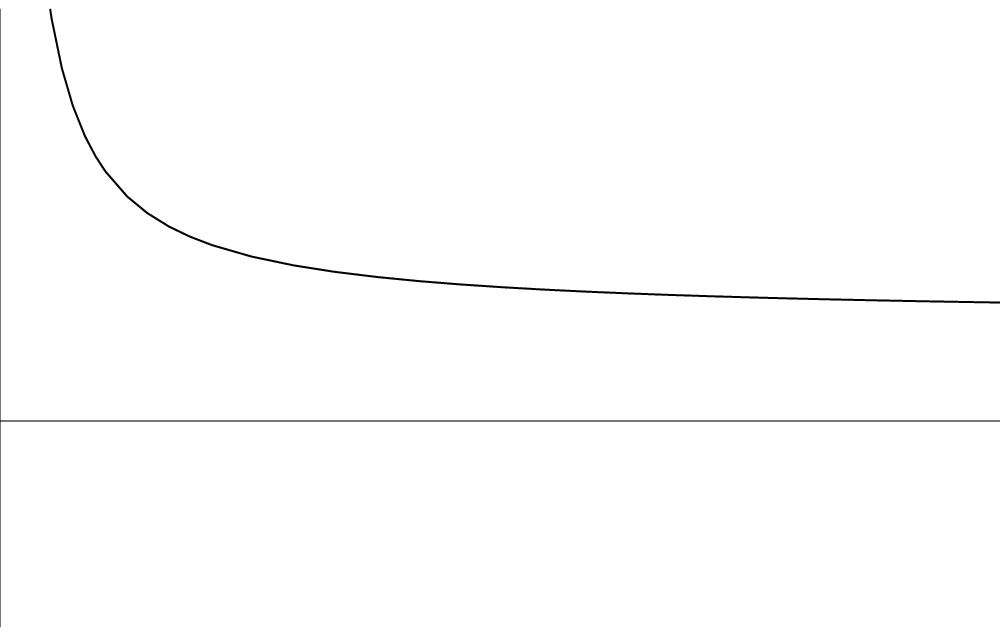,width=1.5in} \
     \ooalign{\raise20pt\hbox{$z$}}  \cr\noalign{\vskip-2\jot}
     \hfil {\Large a)} \hfil & \hfil {\Large b)} \hfil \cr\noalign{\vskip4\jot}
    $V$ & $V$  \cr\noalign{\vskip1\jot}
    \ \psfig{figure=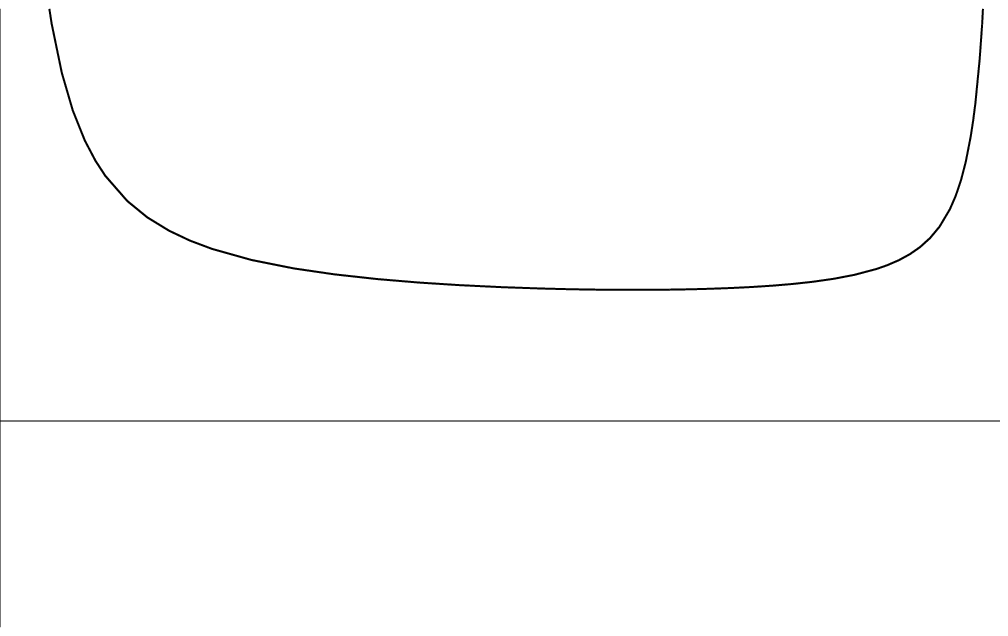,width=1.5in} \
     \ooalign{\raise20pt\hbox{$z$}} &
    \ \psfig{figure=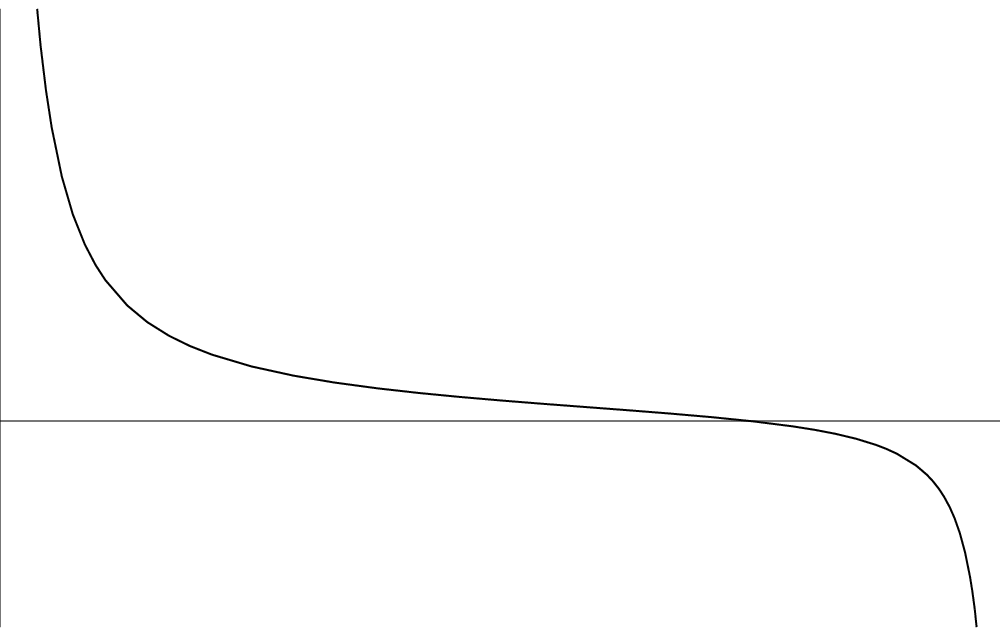,width=1.5in} \
     \ooalign{\raise20pt\hbox{$z$}}  \cr\noalign{\vskip-2\jot}
     \hfil {\Large c)} \hfil & \hfil {\Large d)} \hfil  \cr
  }}$$
 \caption{The various behaviors for $V(z)$ far from the boundary of
$AdS_5$: a) Vanishes; b) Asymptotes to a finite value; c) Increases without
bound; d) Decreases without bound.}\label{figB}
  \end{figure}

Setting
  \eqn{PhiAnsatz}{
   \phi = e^{-i p \cdot x} e^{-3A(z)/2} R(z) \ ,
  }
 one finds that the five-dimensional wave-equation $\square\phi = 0$
reduces to
  \eqn{SchrodingerForm}{
   \left[ -\partial_z^2 + V(z) \right] R = p^2 R \qquad \hbox{where}\qquad
    V(z) = {3 \over 2} A''(z) + {9 \over 4} A'(z)^2 \ .
  }
 As usual we work in $+$$-$$-$$-$$-$ convention.  The potential $V(z)$ 
exhibits four different behaviors, which are illustrated in Figure~\ref{figB}.
 The first is encountered for the $n=1$ flow; the second for the $n=2$
flow; the third for $n=3$; and the fourth for $n=4,5$.  The spectrum of
possible values for $s = p^2$ is determined by the form of $V$: it consists
of discrete points for solutions of \SchrodingerForm\ which are
normalizable and/or a continuum corresponding to solutions which are almost
normalizable in the same sense as plane waves are.  For $n=1$ the spectrum
is continuous, and it covers the whole positive real $s$-axis.  For $n=2$
the spectrum is also continuous, but it covers only the interval
$(\ell^2/L^4,\infty)$ on the $s$-axis: there is a mass gap!  For $n=3,4,5$
the spectrum is discrete and positive, and the lowest eigenvalue for $s$
is on the order $\ell^2/L^4$.  Note that the case $n=5$, though
pathological in ten-dimensional origin, is stable with respect to
fluctuations of a minimally coupled scalar.  It would be interesting
to find some way of perceiving the pathology of ``ghost'' D3-branes
with negative tension and negative charge directly in five dimensions.

The spectrum of \SchrodingerForm\ determines the analyticity properties of
the two-point function
  \eqn{TwoPointPi}{
   \Pi_4(p^2) = \int {d^4 p \over (2\pi)^4} e^{i p \cdot x}
    \langle {\cal O}_4(x) {\cal O}_4(0) \rangle
  }
 in the complex $s$-plane, where again $s = p^2$.  The function $\Pi_4(s)$
is analytic except at the points along the $s$-axis which are included in
the spectrum: points in the discrete spectrum
correspond to poles in $\Pi_4(s)$, and intervals in the continuous spectrum
correspond to cuts.  In principle, $\Pi_4(s)$ can be determined from
solutions to \SchrodingerForm\ which approach a constant as $z \to 0$,
using the prescription of \cite{gkPol,witHolOne}.  In practice one needs an
explicit solution to make much progress, and so far we have results only
for $n=2$ and $n=4$.  To compute the two-point function for $n=2$, the
relevant solutions to the wave equation $\square \phi = 0$ is
  \eqn{HyperSolve}{\eqalign{
   \phi &= e^{-i p \cdot x} v^a F(a,a;2+2a;v) 
      \cr\noalign{\vskip-2\jot}
     & \hbox{where} \quad v = 1/\lambda^6 \quad \hbox{and} \quad 
    a = -{1\over 2} +{1\over 2} \sqrt{1-{L^4 p^2\over \ell^2}}
  }}
 where $F(a,b;c;v)$ is the hypergeometric function.  The two-point function
is 
  \eqn{FinalTwoPoint}{
   \Pi_4(s) = -{N^2 \over 32 \pi^2} s^2 \,\psi\Big({1\over 2}+
   {1\over 2}\sqrt{1-{L^4 s\over \ell^2}}\,\Big) \ ,
  }
 where $\psi(z) = \Gamma'(z)/\Gamma(z)$.  The cut across the real $s$-axis
extends over the interval $(\ell^2/L^4,\infty)$, and this is indeed the
spectrum of \SchrodingerForm.  The discontinuity across the cut is related
to an absorption cross-section where an $s$-wave dilaton falls into the
branes from asymptotically flat infinity (far from the D3-branes).

For $n=4$, the relevant solution to $\square \phi = 0$ is\footnote{An
equation equivalent to $\square \phi = 0$ for $n=4$ arose in the study of
Euclideanized D3-branes with a single large imaginary angular momentum
\cite{Csaki:1998cb,CvGu2}.  This is not a surprise, since the $n=2$ and
$n=4$ metrics are related via $\ell \to i\ell$, and this same replacement
is necessary in Wick rotating $n=2$ to Euclidean signature.  The discrete
``glueball'' spectrum computed there (numerically) coincides with
\eno{WherePoles}, but it appears to have a rather
different interpretation in this context: the Higgs VEV's are responsible
for the energy scale, not confinement.  S.S.G.~would like to thank
M.~Cvetic for useful discussions regarding discrete spectra in similar
contexts.}
  \eqn{PhiSolns}{\eqalign{
   \phi &= e^{-i p \cdot x} F(a,-1-a;1;u) 
      \cr\noalign{\vskip-2\jot}
     & \hbox{where} \quad u = \lambda^6 \quad \hbox{and} \quad 
    a = -{1\over 2} +{1\over 2} \sqrt{1+{L^4 p^2\over \ell^2}} \ .
  }}
 The two-point function, 
  \eqn{TPM}{
   \Pi_4(s) = -{N^2 \over 64 \pi^2} s^2 \left[ 
     \psi\Big({1\over 2} + {1\over 2}\sqrt{1+{L^4 s\over \ell^2}}\,\Big) + 
     \psi\Big({1\over 2} - {1\over 2}\sqrt{1+{L^4 s\over \ell^2}}\,\Big)
    \right] \ ,
  }
 has poles at $s = m^2$ where
  \eqn{WherePoles}{
   m^2 = {4 \ell^2 \over L^4} j(j+1) \qquad \hbox{for} \quad 
    j = 1,2,3,4,\ldots \ .
  }
 These are precisely the excited state energy levels of a rigid rotator,
but we do not see any obvious interpretation of $j$ 
as an angular momentum quantum number.  The
wave functions for these values of $m^2$ are hypergeometric polynomials.

It is straightforward to extend the analysis to two-point functions of
operators corresponding to partial waves of the dilaton whose angular
momenta are in planes with $\theta = 0$.  We will not go into
detail here, but only state that it does not affect the qualitative
behavior of $V$, and for $n=2$ it does not even effect the numerical value
of the gap.  Partial waves with angular momentum not perpendicular to the
D3-branes lead in general to non-separable partial differential equations
in our variables, but we expect the same qualitative conclusions to stand.

In weakly coupled gauge theory, the behavior of the two-point function is
very different.  The operator ${\cal O}_4$ can create two gauge bosons, and
the two-point function at zero 't~Hooft coupling can be evaluated from 
a one-loop graph with two ${\cal O}_4$ insertions.
 In the conformal vacuum of super-Yang-Mills theory, the results
of \cite{IgorAbsorb} indicate that the one-loop 
graph, with only gauge bosons running around the loop,
gives exact agreement with the
strong coupling result.  The subsequent understanding of this agreement
from the point of view of non-renormalization theorems 
\cite{Anselmi:1997am,Anselmi:1998ys,Gubser:1997se,Howe:1998zi,Petkou:1999fv}, 
and the role of lower spin fields and on-shell ambiguities in
${\cal O}_4$, are for us side issues, because none of the
non-renormalization theorems is expected to hold away from the origin of
the moduli space.  The masses of individual gauge bosons or their
superpartners are protected, and this is because they are BPS excitations;
but this does not imply that interactions cannot correct the one-loop
graph.

 \def\mw{\langle m_W \rangle}
 The distribution of masses of gauge bosons follows from the distribution
of branes through the formula
  \eqn{RhoFromSigma}{
   \rho(m) = \alpha' \Vol S^{n-1} (\alpha' m)^{n-1} \int d^n y \, 
    \sigma(\vec{y}) \sigma(\vec{y} + \alpha' m \hat{e}) \ ,
  }
where $\hat{e}$ is an arbitrary unit vector in $n$ dimensions.
 In all cases the maximum mass for a gauge boson is $2\ell/\alpha'$,
because the diameter of $\sigma(\vec{y})$ is $2\ell$.  The average gauge
boson mass, $\mw$, is also $\ell/\alpha'$ up to a factor of order unity.
In \RhoFromSigma\ we have normalized $\rho$ so that $\int dm \, \rho(m) =
1$.  The qualitative features follow from the support of $\rho$, which is
$(0,2\ell/\alpha')$, and the behavior near $m=0$, which is
  \eqn{mZeroRho}{\eqalign{
   n=1: \qquad & \rho(m) \sim {1 \over \mw}  \cr
   n=2: \qquad & \rho(m) \sim {m \over \mw^2}  \cr
   n=3: \qquad & \rho(m) \sim {m^2 \over \mw^3} \log {\mw \over m}  \cr
   n=4: \qquad & \rho(m) \sim {m^2 \over \mw^3} \ . 
  }}
 In the weakly coupled gauge theory, each massive species contributes a
pair-production threshold to the discontinuity in $\Pi_4(s)$.  The total
discontinuity has the approximate form
  \eqn{DiscPiWeak}{
   \Disc \Pi_4(s) \approx N^2 s^2 \int_0^{\sqrt{s}/2} dm \, 
    \rho(m) \sqrt{1 - {4m^2 \over s}} \ .
  }
 One recovers the conformal limit $\Disc \Pi_4(s) \sim N^2 s^2$ for $s \gg
\mw^2$, with corrections suppressed by powers of $\mw^2/s$.  If $\rho(m)
\sim m^\delta/\mw^{1+\delta}$ for $m \ll \mw$, then $\Disc \Pi_4(s) \sim
\sqrt{s}^{5+\delta}/\mw^{1+\delta}$ for $s \ll \mw^2$.  This is in contrast
with the supergravity results: there the conformal limit is recovered for
$s \gg \ell/L^2$, which is a much lower energy since $\ell/L^2 \sim
\mw/\sqrt{g_{YM}^2 N}$; also, in the $n=2$ case, one can show that
corrections to $\Disc \Pi_4(s) \sim N^2 s^2$ are suppressed by powers of
$e^{-\pi L^4 s/\ell^2}$.  Also the gapped spectrum for $n=2$ and the
discrete spectrum for $n=3,4$ are in contrast with the expectation based on
the continuous distribution of branes.  In summary, the two-point function
$\Pi_4(s)$ exhibits nearly conformal power-law behavior down to a much
lower energy scale than the typical gauge boson mass.  Below this low scale
the physics is radically different from gauge theory expectations, and very
sensitive to $n$.  We will come back to this conundrum in
section~\ref{Discussion}.

\section{Wilson loops}
\label{Wilson}

To compute the quark-anti-quark potential from Wilson loops on the
supergravity side, we follow \cite{rWilson,mWilson}.  The asymptotically
$AdS_5$ geometry controls the small $r$ behavior: $V_{q\bar{q}}(r) \sim
\sqrt{g_{YM}^2 N}/r$.  The deviations from the Coulomb law become important
on a length scale $L^2/\ell$, rather than $\alpha'/\ell$ as in the weakly
coupled gauge theory.  Beyond this point, one sees a stronger power law for
$n=1$, and perfect screening (with a caveat which we will come to shortly)
for $n>1$.

Because there is no dilaton profile, the ten-dimensional string metric and
Einstein metric are the same, and ``Wilson'' loops (more properly 't~Hooft
loops) built from D1-branes will show the same behavior as those built from
fundamental strings, up to the overall coefficient of $V_{q\bar{q}}$.  Near
the boundary of $AdS_5$, each end of the string can be constrained to lie
anywhere on $S^5$.  Most of the trajectories do not lie in a plane, and are
difficult to analyze.  The simple cases are where the string stays either
in the hyperplane of ${\bf R}^6$ which contains the brane distribution, or
in the orthogonal hyperplane.  The first case is $\theta = \pi/2$ for $n
\leq 3$ and $\theta = 0$ for $n > 3$, in the coordinate systems used in
\WarpedMetrics; the second case is $\theta = 0$ for $n \leq 3$ and $\theta
= \pi/2$ for $n>3$.  The analysis proceeds most straightforwardly with a
radial variable $u$ such that $\hat{g}_{tt} \hat{g}_{uu} = -1$.  Then the distance
between the quark and anti-quark and the potential energy between them are
  \eqn{qqRV}{\eqalign{
   r &= \int_C dx  \cr
   V_{q\bar{q}} &= {1 \over 2\pi \alpha'} \int_C dx \, \sqrt{f(u) + 
    \left( {\partial u \over \partial x} \right)^2}  \cr
   f(u) &= -\hat{g}_{tt} \hat{g}_{xx} \ ,
  }}\fixit{Make sure of the coefficient on $V$}
 where $C$ is the trajectory of the Wilson loop in the $x$--$u$ plane.  By
assumption, $u=0$ is the location of the branes (in our cases, it is where
curvatures become infinite).  Assuming convex $f(u)$, one can proceed as in
Appendix~A of \cite{GPPZDil} to determine the qualitative behavior of
$V_{q\bar{q}}(r)$.  Namely, if $f(u) \sim u^\gamma$ with $0 < \gamma < 2$,
then there is perfect screening ($V_{q\bar{q}}=0$) at sufficiently large
$r$; if $f(u) \sim u^\gamma$ with $\gamma > 2$, then $V_{q\bar{q}}(r) \sim
r^{2/(2-\gamma)}$ (note that $\gamma = 4$ corresponds to the Coulomb law),
and if $f(u)$ is bounded below, then one obtains an area law
$V_{q\bar{q}}(r) \sim r$.

It is straightforward to transform to variables $u$ in each of the ten
cases we consider, or to derive the general result that if $\hat{g}_{tt} \sim
r^\alpha$ and $\hat{g}_{rr} \sim r^\beta$, then $\gamma =
4\alpha/(2+\alpha+\beta)$.  A subtlety arises when $n>3$: in these cases
the distribution of branes is at $r=\ell$ rather than $r=0$, so one should
replace $r \to r+\ell$ before performing the scaling analysis around $r=0$.
The results are quoted in Table~\ref{tableB}.
  \begin{table}
  \begin{center}
  \begin{tabular}{ccc}
   &$ \theta = 0 $&$ \theta = \pi/2 $  \\[3pt] \hline \\[-7pt]
   $ n=1 $&$ V_{q\bar{q}} \sim 1/r^2 $&$ V_{q\bar{q}} \sim 1/r^4 $  \\
   $ n=2 $& perfect screening & perfect screening  \\
   $ n=3 $& perfect screening & perfect screening  \\
   $ n=4 $& perfect screening & ``confinement''  \\
   $ n=5 $& perfect screening & ``confinement''
  \end{tabular}
  \end{center}
 \caption{Quark/anti-quark interactions derived from Wilson loops
with two different orientations relative to distribution
of branes.}\label{tableB}
  \end{table}
 For the $n=2$, $\theta = 0$ case we find that perfect screening sets
in at $r = \pi L^2/2\ell$.  We have put ``confinement'' in quotations in
Table~\ref{tableB} because it is really a fake: while it is true that a
Wilson loop constrained to lie in the $\theta = \pi/2$ plane for $n>3$ does
exhibit an area law, a physical string at large $r$ would eventually find
it energetically favorable to creep up toward the $\theta = 0$ plane and
enjoy perfect screening.  There is a spontaneous $SO(n)$ symmetry breaking
associated with the orientation of the string in the $(n-1)$-sphere of
\WarpedMetrics.  This is the caveat we mentioned in the first section.

The weak coupling gauge theory expectation, given a distribution $\rho(m)$
which is cut off around $m = \mw$ and has the behavior $\rho(m) \sim
m^\delta / \mw^{1+\delta}$ for $m \ll \mw$, is
  \eqn{VqqWeak}{
   V_{q\bar{q}}(r) = g_{YM}^2 N \int dm \, \rho(m) {e^{-mr} \over r} \sim
    \left\{ \seqalign{\span\TR \quad & \span\TT}{
     {g_{YM}^2 N \over r} & for $r \ll {1 \over \mw}$  \cr
     {g_{YM}^2 N \over \mw^{1+\delta} r^{2+\delta}} & 
      for $r \gg {1 \over \mw}\,$.} \right.
  }
 As before, $\mw = \ell/\alpha'$ up to a factor of order unity.
Interestingly, the infrared power law in the $n=1$ case is $1/r^2$, just as
we saw for $\theta = 0$ in supergravity.

\section{Discussion}
\label{Discussion}

We have constructed and studied geometries which have simple descriptions
both in five-dimensional gauged supergravity as asymptotically $AdS_5$
geometries with profiles for some scalars in $SL(6,{\bf R})/SO(6)$, and in
${\cal N}=4$ super-Yang-Mills theory as vacua on the Coulomb branch.  The
ten-dimensional geometry, composed of $N$ parallel D3-branes in some
continuous distribution in the ${\bf R}^6$ space perpendicular to their
world-volumes, leaves little doubt of the Coulomb branch interpretation;
but the gapped or discrete spectra in the two-point function and the
perfect screening observed in Wilson loops do not seem compatible with
gauge theory expectations.  In particular, there just isn't a mass gap of
size $\ell/L^2$ in the gauge theory: one can construct color singlet states
of lower mass by putting two light gauge bosons far apart.

Before suggesting a possible resolution, let us re-examine the energy
scales involved.  A typical gauge boson has mass $\mw = \ell/\alpha'$, so
this is the energy scale at which one would expect deviations from
conformality to become important.  But the two-point function and Wilson
loop calculations identify the much smaller energy ${\cal E}_c = \ell/L^2
= \mw/\sqrt{g_{YM}^2 N}$ as the scale at which conformal invariance is
substantially lost and the interesting dynamics (e.g.\ screening, gaps, and
discrete spectra) takes place.  In a sense this is precisely the
discrepancy in normalization of energy scales observed in \cite{PeetPolch}:
when converting an energy into a value of the radial coordinate $U$,
energies such as $\mw$ pertaining to stretched strings differ by a factor
of $\sqrt{g_{YM}^2 N}$ from the conversion appropriate to supergravity
probes.  In the present context, $U$ can be generalized to a coordinate
system $U_i = y_i/\alpha'$ on the ${\bf R}^6$ perpendicular to the branes.
This does not seem a satisfactory resolution because a mass gap in an
absorption calculation is something that can be compared to masses of brane
excitations without any $\sqrt{g_{YM}^2 N}$ ambiguity.

A feature that all our geometries share is that curvatures become large
close to the brane distribution.  This raises the possibility that an
analog of the Horowitz-Polchinski correspondence principle \cite{HoroPolch}
is at work.  For specificity let us consider only the $n=2$ case.  There is
a ``halo,'' of thickness $\ell/\sqrt{g_{YM}^2 N}$ in the flat metric
$\sum_i dy_i^2$, surrounding the disk of branes in ${\bf R}^6$, inside
which curvatures are stringy.  Outside this halo supergravity applies, and
it keeps track of the strong coupling super-Yang-Mills dynamics at high
energies; inside, or at lower energies, one may expect that a direct gauge
theory description becomes practical.  An open string running from the disk
to a test D3-brane on the edge of the halo has a mass on the order
$\ell/L^2$.  At this energy scale, one may argue that the gauge theory is
no longer strongly coupled because most of the degrees of freedom have been
integrated out: the large $N$ in the 't~Hooft coupling $g_{YM}^2 N$ is
substantially reduced.

In this picture, a natural expectation would be that the gap, the discrete
spectrum, and perfect screening will all be washed out in the process of
matching onto the low-energy weakly coupled gauge theory description, to be
replaced with the power law behaviors we described at the ends of
sections~\ref{TwoPoint} and~\ref{Wilson}.  This indeed is one possible
resolution of our difficulties.  It is not a complete resolution because
there is still the region of energies $\ell/L^2 \ll {\cal E} \ll
\ell/\alpha'$ where supergravity is valid and gives nearly conformal
predictions at odds with the Higgs mass scale in the gauge theory.  If it
is taken seriously, then the results of \cite{minDil,gDil,GPPZDil}
regarding confinement from similarly singular supergravity geometries must
be regarded as suspect.  However it seems possible to argue, both in our
case and in \cite{minDil,gDil,GPPZDil}, that in an appropriate large $N$,
large $g_{YM}^2 N$ limit, the wave-function overlap of energy
eigenfunctions with the region of stringy or Planckian curvatures is
controllably small.  In such a limit the most one would expect is a slight
broadening of the eigen-energy delta functions.

As another possible resolution, we would like to suggest that the gauge theory
physics might not be as featureless as the usual Coulomb branch analysis
implies.  The supergravity solution specifies only a continuous
distribution of branes, $\sigma(\vec{w})$, which can only be approximated
by the $N$ branes at our disposal.  It seems more natural to regard
$\sigma(\vec{w})$ as specifying not a single distribution of branes, but an
ensemble of distributions where the $N$ branes are allowed to move slightly
relative to one another.  One should then include an integration over the
ensemble in the path integral: rather taking a specific point in moduli
space as the vacuum, one should integrate over the region of moduli space
which is consistent with the distribution $\sigma(\vec{w})$.  If this
integration is done first, its effect is to induce extra interaction terms
in the lagrangian.  With regard to color indices, these terms do not have a
pure trace structure.  Keeping only the lowest dimension operators, the
schematic form we expect for the lagrangian is
  \eqn{LowDim}{
   {\cal L}_{\rm eff} = \tr (\partial X_i)^2 + \tr [X_i,X_j]^2 + \lambda 
    ({\cal O}_{20'})^2 + 
    \ldots \ ,
  }
 where for simplicity we keep only the scalar fields and work in Euclidean
signature.  The operator ${\cal O}_{20'}$ is the dimension two $\tr X_{(i}
X_{j)}$ operator whose VEV characterizes the state.\fixit{notation on
$SO(6)$ indices.}  The $SO(6)$ singlet operator $\tr \sum_i X_i^2$ is excluded on
the grounds that AdS/CFT predicts a large dimension for it
\cite{gkPol,witHolOne}.  The size of $\lambda$ is controlled by how densely the
branes are packed in the distributions approximating $\sigma(\vec{w})$: the
sparser the distribution, the larger is $\lambda$.  

\begin{figure}
  $$\seqalign{\span\TL \quad & \span\TC \quad & \span\TR & 
   \span\TC & \span\TC \quad & \span\TR}{
   \langle {\cal O}_{20'} \rangle \hskip-30pt \span = \span 
   \hskip-30pt
   \ooalign{\lower30pt\hbox{\psfig{figure=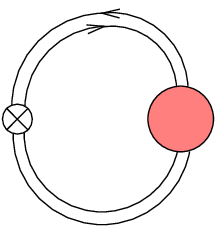}}}  
    \cr\noalign{\vskip-18\jot}
   \ooalign{\lower9pt\hbox{\psfig{figure=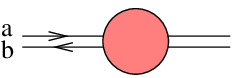}}} &=& 
   \ \psfig{figure=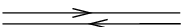} &\mp&
   \ooalign{\lower6pt\hbox{\psfig{figure=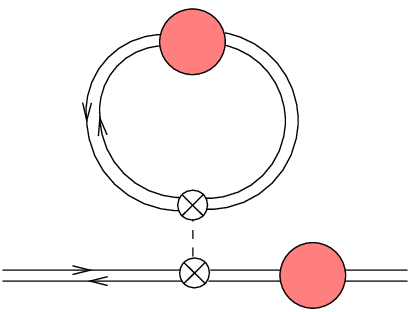}}} & 
     + \ldots  \cr\noalign{\vskip3\jot}
   {1 \over p^2 + m_{ab}^2} &=& \ {1 \over p^2 + m_{(0)ab}^2} &\mp&
    {\lambda \langle {\cal O}_{20'} \rangle \over 
    (p^2 + m_{(0)ab}^2)(p^2 + m_{ab}^2)} & + \ldots  \cr\noalign{\vskip3\jot}
   \hfill m_{ab}^2 &=& 
     m_{(0)ab}^2 \pm \lambda \langle {\cal O}_{20'} \rangle + \ldots \hfill \span 
  }$$
 \caption{Self-consistent treatment of leading order color-independent
corrections to masses of the scalar $X_i$.  The shaded circle indicates the
full dressed propagator, and $a$ and $b$ are color indices.  The operator
$({\cal O}_{20'})^2$ is represented as a dotted line connecting the two
${\cal O}_{20'}$ insertions.  The $\pm$ sign is chosen according as ${\cal
O}_{20'}$ includes a positive or negative mass term for $X_i$.}\label{figD}
  \end{figure}

The double trace form of $({\cal O}_{20'})^2$ leads to color-independent
mass corrections through diagrams shown schematically in
figure~\ref{figD}.\footnote{We thank D.~Kabat for a useful discussion
regarding the use of a similar mechanism in another context \cite{Kabat}.}
Typically one expects such mass corrections to be negative because they
come from a second order effect in perturbation theory, but because ${\cal
O}_{20'}$ is a traceless combination of mass terms for the scalars, at
least some of the mass corrections are positive.  Also, bubble graphs built
using $({\cal O}_{20'})^2$ contribute corrections to the two-point function
$\Pi_4(s)$.  It is possible that if the mass corrections or the
interactions due to $({\cal O}_{20'})^2$ are large, they may change the
physics enough to induce the mass gap at $\ell/L^2$, the discrete spectra,
and/or the screening observed in sections~\ref{TwoPoint} and~\ref{Wilson}.
We emphasize the speculative nature of this scenario.  However, the ${\cal
N}=4$ gauge dynamics alone does not seem likely to encompass the variety of
physical behaviors that we have seen, and we take it as a clue that the
deviations from the expected weak coupling behavior become more radical as
the branes become more sparsely distributed.

\section*{Acknowledgements}

We would like to thank M.~Grisaru, E.~Martinec, H.~Saleur, L.~Susskind,
E.~Witten, and particularly J.~Polchinski for useful discussions and
commentary.  In communications with K.~Sfetsos, we have learned that he has
independently obtained results which have some overlap with the present
work.\footnote{Note added: These results have subsequently appeared in
\cite{ABKS}.}

The research of D.Z.F.\ was supported in part by the NSF under grant number
PHY-97-22072. The research of S.S.G.\ was supported by the Harvard Society
of Fellows, and also in part by the NSF under grant number PHY-98-02709,
and by DOE grant DE-FGO2-91ER40654.  The work of K.P.\ and N.W.\ was
supported in part by funds provided by the DOE under grant number
DE-FG03-84ER-40168.


\bibliography{sum}

\begingroup\raggedright\begin{thebibliography}{10}

\bibitem{juanAdS}
J.~Maldacena, ``The Large N limit of superconformal field theories and
  supergravity,'' {\em Adv. Theor. Math. Phys.} {\bf 2} (1998) 231,
  \href{http://xxx.lanl.gov/abs/hep-th/9711200}{{\tt hep-th/9711200}}.

\bibitem{gkPol}
S.~S. Gubser, I.~R. Klebanov, and A.~M. Polyakov, ``Gauge theory correlators
  from noncritical string theory,'' {\em Phys. Lett.} {\bf B428} (1998) 105,
  \href{http://xxx.lanl.gov/abs/hep-th/9802109}{{\tt hep-th/9802109}}.

\bibitem{witHolOne}
E.~Witten, ``Anti-de Sitter space and holography,'' {\em Adv. Theor. Math.
  Phys.} {\bf 2} (1998) 253, \href{http://xxx.lanl.gov/abs/hep-th/9802150}{{\tt
  hep-th/9802150}}.

\bibitem{JMNW}
J.~A. Minahan and N.~P. Warner, ``Quark potentials in the Higgs phase of large
  N supersymmetric Yang-Mills theories,'' {\em JHEP} {\bf 06} (1998) 005,
  \href{http://xxx.lanl.gov/abs/hep-th/9805104}{{\tt hep-th/9805104}}.

\bibitem{9905104}
I.~R. Klebanov and E.~Witten, ``AdS / CFT correspondence and symmetry
  breaking,'' \href{http://xxx.lanl.gov/abs/hep-th/9905104}{{\tt
  hep-th/9905104}}.

\bibitem{fgpw}
D.~Z. Freedman, S.~S. Gubser, K.~Pilch, and N.~P. Warner, ``Renormalization
  group flows from holography---supersymmetry and a c-theorem,''
  \href{http://xxx.lanl.gov/abs/hep-th/9904017}{{\tt hep-th/9904017}}.

\bibitem{BdWHN}
B.~de~Wit and H.~Nicolai, ``The consistency of the $S^7$ truncation in $d = 11$
  supergravity,'' {\em Nucl. Phys.} {\bf B281} (1987) 211.

\bibitem{stny}
H.~Nastase, D.~Vaman, and P.~van Nieuwenhuizen, ``Consistent nonlinear K K
  reduction of 11-d supergravity on AdS(7) x S(4) and selfduality in odd
  dimensions,'' \href{http://xxx.lanl.gov/abs/hep-th/9905075}{{\tt
  hep-th/9905075}}.

\bibitem{GRW}
M.~Gunaydin, L.~J. Romans, and N.~P. Warner, ``Gauged $N=8$ supergravity in
  five-dimensions,'' {\em Phys. Lett.} {\bf 154B} (1985) 268.

\bibitem{PPvN}
M.~Pernici, K.~Pilch, and P.~van Nieuwenhuizen, ``Gauged N=8 D=5
  supergravity,'' {\em Nucl. Phys.} {\bf B259} (1985) 460.

\bibitem{GRWb}
M.~Gunaydin, L.~J. Romans, and N.~P. Warner, ``Compact and noncompact gauged
  supergravity theories in five-dimensions,'' {\em Nucl. Phys.} {\bf B272}
  (1986) 598.

\bibitem{kpw}
A.~Khavaev, K.~Pilch, and N.~P. Warner, ``New vacua of gauged N=8 supergravity
  in five-dimensions,'' \href{http://xxx.lanl.gov/abs/hep-th/9812035}{{\tt
  hep-th/9812035}}.

\bibitem{dz}
J.~Distler and F.~Zamora, ``Nonsupersymmetric conformal field theories from
  stable anti-de Sitter spaces,''
  \href{http://xxx.lanl.gov/abs/hep-th/9810206}{{\tt hep-th/9810206}}.

\bibitem{klt}
P.~Kraus, F.~Larsen, and S.~P. Trivedi, ``The Coulomb branch of gauge theory
  from rotating branes,'' {\em JHEP} {\bf 03} (1999) 003,
  \href{http://xxx.lanl.gov/abs/hep-th/9811120}{{\tt hep-th/9811120}}.

\bibitem{Myers}
A.~Chamblin, R.~Emparan, C.~V. Johnson, and R.~C. Myers, ``Charged AdS black
  holes and catastrophic holography,''
  \href{http://xxx.lanl.gov/abs/hep-th/9902170}{{\tt hep-th/9902170}}.

\bibitem{CvGu1}
M.~Cvetic and S.~S. Gubser, ``Phases of R charged black holes, spinning branes
  and strongly coupled gauge theories,''
  \href{http://xxx.lanl.gov/abs/hep-th/9902195}{{\tt hep-th/9902195}}.

\bibitem{PopeEtAl}
M.~Cvetic {\em et.~al.}, ``Embedding AdS black holes in ten-dimensions and
  eleven-dimensions,'' \href{http://xxx.lanl.gov/abs/hep-th/9903214}{{\tt
  hep-th/9903214}}.

\bibitem{bcs}
K.~Behrndt, M.~Cvetic, and W.~Sabra, ``Nonextreme black holes of
  five-dimensional N=2 AdS supergravity,''
  \href{http://xxx.lanl.gov/abs/hep-th/9810227}{{\tt hep-th/9810227}}.

\bibitem{ksDil}
A.~Kehagias and K.~Sfetsos, ``On asymptotic freedom and confinement from type
  IIB supergravity,'' \href{http://xxx.lanl.gov/abs/hep-th/9903109}{{\tt
  hep-th/9903109}}.

\bibitem{minDil}
J.~A. Minahan, ``Asymptotic freedom and confinement from type 0 string
  theory,'' \href{http://xxx.lanl.gov/abs/hep-th/9902074}{{\tt
  hep-th/9902074}}.

\bibitem{gDil}
S.~S. Gubser, ``Dilaton driven confinement,''
  \href{http://xxx.lanl.gov/abs/hep-th/9902155}{{\tt hep-th/9902155}}.

\bibitem{GPPZDil}
L.~Girardello, M.~Petrini, M.~Porrati, and A.~Zaffaroni, ``Confinement and
  condensates without fine tuning in supergravity duals of gauge theories,''
  \href{http://xxx.lanl.gov/abs/hep-th/9903026}{{\tt hep-th/9903026}}.

\bibitem{Csaki:1998cb}
C.~Csaki, Y.~Oz, J.~Russo, and J.~Terning, ``Large N QCD from rotating
  branes,'' {\em Phys. Rev.} {\bf D59} (1999) 065012,
  \href{http://xxx.lanl.gov/abs/hep-th/9810186}{{\tt hep-th/9810186}}.

\bibitem{CvGu2}
M.~Cvetic and S.~S. Gubser, ``Thermodynamic stability and phases of general
  spinning branes,'' \href{http://xxx.lanl.gov/abs/hep-th/9903132}{{\tt
  hep-th/9903132}}.

\bibitem{IgorAbsorb}
I.~R. Klebanov, ``World volume approach to absorption by nondilatonic branes,''
  {\em Nucl. Phys.} {\bf B496} (1997) 231,
  \href{http://xxx.lanl.gov/abs/hep-th/9702076}{{\tt hep-th/9702076}}.

\bibitem{Anselmi:1997am}
D.~Anselmi, D.~Z. Freedman, M.~T. Grisaru, and A.~A. Johansen,
  ``Nonperturbative formulas for central functions of supersymmetric gauge
  theories,'' {\em Nucl. Phys.} {\bf B526} (1998) 543,
  \href{http://xxx.lanl.gov/abs/hep-th/9708042}{{\tt hep-th/9708042}}.

\bibitem{Anselmi:1998ys}
D.~Anselmi, J.~Erlich, D.~Z. Freedman, and A.~A. Johansen, ``Positivity
  constraints on anomalies in supersymmetric gauge theories,'' {\em Phys. Rev.}
  {\bf D57} (1998) 7570--7588,
  \href{http://xxx.lanl.gov/abs/hep-th/9711035}{{\tt hep-th/9711035}}.

\bibitem{Gubser:1997se}
S.~S. Gubser and I.~R. Klebanov, ``Absorption by branes and Schwinger terms in
  the world volume theory,'' {\em Phys. Lett.} {\bf B413} (1997) 41--48,
  \href{http://xxx.lanl.gov/abs/hep-th/9708005}{{\tt hep-th/9708005}}.

\bibitem{Howe:1998zi}
P.~S. Howe, E.~Sokatchev, and P.~C. West, ``Three point functions in N=4
  Yang-Mills,'' {\em Phys. Lett.} {\bf B444} (1998) 341,
  \href{http://xxx.lanl.gov/abs/hep-th/9808162}{{\tt hep-th/9808162}}.

\bibitem{Petkou:1999fv}
A.~Petkou and K.~Skenderis, ``A Nonrenormalization theorem for conformal
  anomalies,'' \href{http://xxx.lanl.gov/abs/hep-th/9906030}{{\tt
  hep-th/9906030}}.

\bibitem{rWilson}
S.-J. Rey and J.~Yee, ``Macroscopic strings as heavy quarks in large N gauge
  theory and anti-de Sitter supergravity,''
  \href{http://xxx.lanl.gov/abs/hep-th/9803001}{{\tt hep-th/9803001}}.

\bibitem{mWilson}
J.~Maldacena, ``Wilson loops in large N field theories,'' {\em Phys. Rev.
  Lett.} {\bf 80} (1998) 4859,
  \href{http://xxx.lanl.gov/abs/hep-th/9803002}{{\tt hep-th/9803002}}.

\bibitem{PeetPolch}
A.~W. Peet and J.~Polchinski, ``UV / IR relations in AdS dynamics,'' {\em Phys.
  Rev.} {\bf D59} (1999) 065011,
  \href{http://xxx.lanl.gov/abs/hep-th/9809022}{{\tt hep-th/9809022}}.

\bibitem{HoroPolch}
G.~T. Horowitz and J.~Polchinski, ``A Correspondence principle for black holes
  and strings,'' {\em Phys. Rev.} {\bf D55} (1997) 6189--6197,
  \href{http://xxx.lanl.gov/abs/hep-th/9612146}{{\tt hep-th/9612146}}.

\bibitem{Kabat}
D.~Kabat, talk on joint work with G.~Lifschytz at ``Black Holes II,'' Montreal,
  June 1999.

\bibitem{ABKS}
A.~Brandhuber and K.~Sfetsos, ``Wilson loops from multicentre and rotating
  branes, mass gaps and phase structure in gauge theories,''
  \href{http://xxx.lanl.gov/abs/hep-th/9906201}{{\tt hep-th/9906201}}.

\end{thebibliography}\endgroup
\bibliographystyle{ssg}

\end{document}